\documentclass[11pt]{article}
\usepackage[utf8]{inputenc}
\usepackage{hyperref}
\usepackage{amsmath}
\usepackage{color}
\usepackage{graphicx}
\usepackage{enumerate}
\usepackage[top=1in, bottom=1in, left=1in, right=1in]{geometry}

\title{Epidemic spread, parameter sensitivity and vaccination strategies on a  random graph with overlapping communities}
\author{\'Agnes Backhausz$^{1,2}$ \ and Gy\"orgy J.\ Sz\'ekely$^1$}

\date{\today}

\begin{document}

\maketitle

\begin{abstract}
Our main goal is to examine the role of communities in epidemic spread in a random graph model. More precisely, we consider a random graph model which consists of overlapping complete graphs, representing households, workplaces, school classes, and which also has a simple geometric structure. We study the model's sensitivity to infection parameters and other tunable parameters of the model, which might be helpful in finding efficient social distancing strategies. We also quantitatively compare different vaccination strategies to see which order is the best to defend the most vulnerable groups or the population in general, and how important it is to gather and use information on the position of infected individuals in the network.
\end{abstract}

\noindent {\bf MSc classification:} 92D30


\footnotetext[1]{ELTE E\"otv\"os Lor\'and University, Budapest, Hungary, Faculty of Science, Institute of Mathematics, P\'azm\'any P\'eter s\'et\'any 1/c, Budapest, Hungary, H-1117}

\footnotetext[2]{Alfr\'ed R\'enyi Institute of Mathematics, Re\'altanoda utca 13-15., Budapest, Hungary, H-1053}

\section{Introduction}

The society we live in mainly consists of smaller or larger communities like households, workplaces, and schools. From the point of view of spreading processes, this overlapping community structure might play a key role. For example, in case of epidemic spread, an infected adult can transmit the disease to his colleagues, who can infect their children, who can then bring the infection to their school classes, and so in. This also means that we can think of infection not as a separate event, but as a potential risk to every member of that community. Hence, when we plan to control a spreading process (for example, to prevent an epidemic wave by social distancing or vaccination), it might be useful to handle the members of the same household, workplace, school class, etc. together. This approach is also motivated by the work of Bok\'anyi, Heemskerk, and Takes \cite[2023]{bokanyi2023anatomy}, who examined family, work, and school connections obtained from administrative registers, and studied this structure from the point of view of network analysis (e.g.\ degree distribution, closure, distances between the different nodes). From a slightly different point of view, the paper of  Cherifi, Palla,  Szymanski, and Lu \cite[2019]{cherifi2019community} presents several random graph models that are based on dense communities, and analyses the role of this in epidemic spread. However, in their models, such as the stochastic block models, these communities are typically not overlapping.

In the current paper our main goal is to study the effects of the community structure in a four-layer random graph model, which is a refined and more realistic version of the model of \cite{fourlayer}. In our setup, this means a network consisting of four layers, representing households, workplaces (together with schools and nursing homes), geometric structures, and public areas. Individuals belong to one of the three age groups. The graph's first two layers have a very simple structure: we randomly split the individuals into disjoint groups (according to a uniform distribution given the age groups), and connect each pair of individuals belonging to the same group. Since the random construction of the cliques in the two layers are independent, and the age distribution is also different (households have a more inhomogeneous age profile in our model than workplaces or schools), we already obtain a nontrivial structure from the point of view of epidemic spread. So far, our model can be viewed as an extended version of the so-called household model, which was first studied by Ball and Scalia-Tomba \cite[1997]{ball}. In this model, households are connected to each other with a complete graph, without any network or age structure. This model has been studied from various points of view, e.g.\ optimal vaccination strategies \cite[Ball--Sirl, 2018]{ballvacc}; see also a survey on this topic \cite[Ball, Britton, House, Isham, Mollison, Pellis, Scalia-Tomba, 2015]{ball2015seven}.  However, as we can learn for example from studies like the paper of Aleta, Martin-Corral, and Pastore \cite[2020]{aleta2020pionttia}, where real data from smartphone applications were used to track people's movements and contacts, the underlying geometric structure can also have a very important role in epidemic spread. Hence by adding a third layer to our model, representing the location where the members of each family live, and a fourth one, representing the public areas where people living close to each other meet, we get an even more realistic model. This way, the household layer with the geometric structure becomes somewhat similar to the spatial modular model of Gross and Havlin \cite[2020]{gross2020epidemic}; however, in their case, there are much larger communities at the nodes of the grid, which themselves have a more complex network structure.  More generally, with the small complete graphs connected to each other with randomly chosen edges, our model has common elements with the model of Stegehuis, van der Hofstad and van Leeuwaarden \cite[2016]{stegehuis2016epidemic}, where, starting from real-world networks, the edges between the smaller communities were rewired randomly.  Shang, Liu, Xie, and Wu \cite[2014]{shang2014overlapping} also studied networks with overlapping communities, and examined the role of vertices belonging to more than one group. As for controlling the epidemic,  Gupta, Singh, and Cherifi \cite[2015]{gupta2015community} showed that a method based on information about smaller communities in the network might be significantly better than strategies using centrality or other local information. This also suggests that it is interesting to see whether vaccinating the members of households, school classes, or workplaces at the same time helps reduce the size of the epidemic wave in our model. 

Although our model has a fixed structure, there are many parameters that we can tune. Since we mainly focus on the effects of the community structure of the graph, we do not choose the parameters to fit any real epidemic curves (especially since these observations are often affected by various incidental events, e.g.\ lack of testing capacity or lagging in reporting death or recovery).  Instead, we have chosen the parameters such that the frequency of contacts approximately matches observations of a representative survey in Hungary (Koltai, V\'as\'arhelyi, R\"ost and Karsai, \cite[2022]{maszk}). Combining these with data from a census in Hungary and observations from the Covid-19 pandemic in Hungary, we can obtain a realistic choice of parameters. Then, with the help of computer simulations, we can compare the effectiveness of social distancing strategies (e.g.\ reducing the number of classes in schools or closing certain workplaces) or vaccination strategies. In particular, we will compare strategies that do not use the community structure of the graph with other ones that are heavily based on this, for example, when we vaccinate members of a school class or workplace. This helps us find effective measurements and also gives us hints about what kind of information can be useful to be collected if we want to control epidemic spread.  

In \autoref{sec:model} we present the four-layer random graph model, together with the data sets that we will use. In \autoref{sec:epidemic}, we sketch the epidemic model that we use and check that the parameters that we use provide a realistic setup. This also includes a heuristic argument for calculating the basic reproduction number $R_0$. In  \autoref{sec:vaccine}, we compare different vaccination strategies, depending on age, network structure, or both. These studies are useful to decide whether it is worth collecting information about the community structure of the population and making an effort to organize vaccination such that members of the same groups get immunization at the same time.

\section{Constructing a four-layer random graph model} 

\label{sec:model}

\subsection{Data sets}

\label{sec:data}

Although our main goal is to understand the effect of the different features of the multilayer random graph model and not to fit the model perfectly to reality, we used two data sets to obtain a reasonable parameter setup.  

The first data set that we will use is the distribution of the household configurations from the 2011 census in Hungary \footnote{Hungarian Central Statistical Office, Data of the population census held in 2011.  \url{http://www.ksh.hu/nepszamlalas/docs/tablak/lakas/06_01_02_14.xls}}. More precisely, in our model, the population is divided into three age groups: 0-29 years (group 1, young people), 30-59 years (group 2, adults), and 60+ years (group 3, elderly people). Hence, there are many possible household configurations, for example, two people from age group 1 and two from age group 2, or one single person from age group 3. The observed proportion of each configuration can be found in the census data. For the sake of simplicity, we omitted the least frequent possibilities and used the proportions given by \autoref{tab:household}. 

\begin{table}
\centering
\begin{tabular}{cccc}
$0-29$ & $30-59$ & $60-$ & proportion \\ \hline 
0 & 1 & 0 & $17\%$\\
0 & 2 &0 & $12\%$ \\ 
0 & 3 & 0 & $4\%$ \\
0 & 0 & 1 &  $12\%$\\
0 & 0 & 2 & $7\%$\\
1 & 1 & 0 & $4\%$\\
1 & 2 & 0 & $9\%$\\
2 & 2 & 0 & $12\%$\\
3 & 2 & 0& $5\%$ \\
4 & 2 & 0 & $2\%$ \\
1 & 0 & 1 & $1\%$ \\
0 & 1 & 1 & $7\%$ \\
0 & 2 & 1 & $4\%$ \\
1 & 1 & 1 & $1\%$ \\
1 & 2 & 1 & $2\%$ \\
2 & 2& 1 & $1\%$
\end{tabular}
\caption{Proportion of different household configurations in Hungary, based on the 2011 census; the average household size is $2.43$. The proportion of the different age groups is $27.2\%$, $55.5\%$ and $17.3\%$}
\label{tab:household}
\end{table}

To make the distribution of connections between different households more realistic, we used the contact matrices from the work of Koltai, V\'as\'arhelyi, R\" ost, and Karsai  \cite{maszk}. In March 2020, a  representative survey was performed in Hungary, in which  people were asked on different channels about how many  contacts they had had on a typical day from the different age groups before the pandemic had started. In this survey, a contact person is someone whom we spend at least 15 minutes within a distance of at most 2 meters. Although this is not the only way how airborne viruses spread, we will also use this definition. In \cite{maszk}, there were 9 age groups, so we unified some of these to meet the age groups from the census data (groups $1, 2, 3$ as defined above). We got the following contact matrices, where the  $i$th row corresponds to  age group $i$, and the $j$th element of the row is the daily average number of contacts  of a person from age group $i$ from age group $j$.  (Notice that since the sizes of the age groups are different, the matrix is not symmetric).

\begin{equation}\label{eq:contactc}
    C=\left(\begin{array}{ccc} 7.4 & 4.1 & 1.2 \\ 6.6 &  9  & 3.7 \\ 0.9 & 2.1 & 2.3\end{array}\right)
\end{equation}

For example, for a person from age group $1$, the total number of contacts on average on a day is $12.7$, and, in particular, the average number of contacts from age group $2$ is $4.1$.

We will also need the average number of contacts outside the household where the person lives. By combining the distribution of the household configurations from \autoref{tab:household} and the contact matrix $C$ in equation \eqref{eq:contactc}, we get the following matrix:

\begin{equation}\label{eq:contacto}
    C_{\rm o}=\left(\begin{array}{ccc} 6.2 & 2.2 & 1.1 \\ 5.7 & 8.1   & 3.5  \\ 0.8 & 1.6 & 2.0\end{array}\right)
\end{equation}
Here, the $j$th element of the $i$th row represents the average number of contacts of a person from age group $i$ with people from age group $j$ outside his or her own household. For example, a person from age group $1$ has $9.5$ contacts on average outside his or her own household in total and $2.2$ on average from age group $2$.

\subsection{The multilayer random graph model}

\label{ref:graph}

Our random graph model consists of four layers, as follows. Notice that for the epidemic spread, we will use a weighted version of this graph, that is, the infection rate will be different in the different layers and communities. The model is based on \cite{fourlayer}. However, compared to that model, we fitted the size and structure of the small communities such that the number of contacts is much closer to the contact matrix $C_{\rm o}$ (recall equation \eqref{eq:contacto}), so this version is more realistic. We will also check some important characteristics such as the basic reproduction number $R_0$ later in this section. 

\subsubsection*{Layer 1: households}

Starting from the idea of the household model \cite{ball}, our first layer consists of small disjoint cliques (complete graphs) representing very close contacts. In particular, given the number of households, $N$, we randomize $N$ household configurations independently, according to the distribution from the census data described in \autoref{tab:household}. Then we form disjoint cliques by connecting the vertices representing people living in the same household (that is, each household member is connected to every other with an edge). 

\subsubsection*{Layer 2: schools, workplaces, nursing homes}

In the second layer, for each age group of the population, we constructed a family of small cliques, representing the people who he or she meets every day (or at least on every working day) outside his or her own household. The sizes of these groups were chosen such that the number of contacts is approximately equal to the real data in the matrix $C_0$ (recall equation \eqref{eq:contacto}). 

For people belonging to age group $1$, we constructed the layer of schools. There is already a simplification here, as age group $1$ is for people of age $0-30$ (so that we have appropriate data from the census), and it is not true that all people of this age go to schools. However, younger people often work in large office buildings or other workplaces which have somewhat similar structures, so we will simply call these places schools. On the other hand, we assume that schools are not just simple cliques. In particular, infection is possible between any two young people  who attend the same school, but it is much more likely between students who are in the same class. In our model, all schools have $s_{\rm sch}=200$ students (but if the total number of young people is not divisible by $200$, then there is a smaller school as well), and these are formed uniformly at random from the individuals in age group $1$ (every individual belongs to exactly one school). Then, within each school, we form classes of size $s_{\rm cls}=8$, again uniformly at random,  such that each possible configuration has the same probability. Of course, typical school classes are much larger than size $8$, but since the contact matrix in equation \eqref{eq:contacto} tells us that, on average, the number of contacts of an individual from age group $1$ with individuals from the same age group is $6.2$, eight is a more realistic choice. Furthermore, to represent the contacts between the members of different age groups that we see in $C_0$, for each class, we add a randomly chosen person from age group $2$, and a randomly chosen person from age group $3$, and think of them as teachers. Adults and elderly people belonging to different classes are connected to all other members of the school as well, but, as we will see, with a smaller weight than the weight of the edges within a class. 

Workplaces are formed similarly, with a uniform random choice from the individuals in age group $2$. Given the size of the workplaces, $s_{\rm wp}=10$ in the basic scenario, we form groups of size $s_{\rm wp}$ such that each possible configuration has the same probability. People belonging to the same group are connected with an edge, so we obtain cliques of size $s_{\rm wp}$.  The size of $s_{\rm wp}$ is a realistic choice given the middle element of $C_0$: a person from age $2$ meets $8.1$ other people from the same age group on average on a day. 

The contact matrix $C_0$ tells us that people above the age of $60$ years (age group $3$) on average meet $2$ other people from the same age group on a day. A part of this might come from the fact that many people from this group work (the general age of retirement is $65$ years for men in Hungary), and other people might live in nursing homes. In order to model this kind of contact, we consider the people in age group $3$ who are not counted as teachers in the school, and  we choose $r_{\rm wp}=80\%$ of them. Once we have selected these people,  we form groups uniformly at random and connect the individuals who belong to the same group. This is considered as a part of Layer 2, hence the size of these groups is $s_{\rm wp}=10$, the same as for other working places.  In addition, we also add a person from age group $2$ to these cliques, representing the staff of the nursing home or a younger employee at a workplace. This corresponds to $C_0$ again: we see that, on average, a person from age group $2$ has $3.5$ contacts from age group $3$, while a person from age group $3$ has only $1.6$ contacts from age group $2$ -- this asymmetry is represented by the groups where the majority of the people is from age group $3$. 

\subsubsection*{Layer 3: geometric structure}

This layer is basically the same as in \cite{fourlayer}, so we summarize this only briefly. Given the number of households, $N$, we form a square grid with $N$ vertices, with a denser part in the middle, where we draw the diagonals of the squares as well (we can imagine a block of houses with more connections, and suburbs around it). The households constructed in Layer 1 are assigned to the nodes of the grid (this is the same as assigning a household chosen randomly independently from the household distribution given in  \autoref{tab:household} to each vertex). Then, if two people live in different households that are connected with an edge in this graph, we connect these individuals with an edge as well.      
    
\subsubsection*{Layer 4: public areas}

This is also the same as in \cite{fourlayer}. In order to represent stores, public transport, and other places where many people living close to each other meet, we form groups of the same size $s_{\rm sc}=200$. This partition is not random; it is based on the geometric structure of Layer $3$. We can imagine that each such place (let us call it a store) has a fixed location, and people living within a given distance go there regularly. However, in our model, people do not infect each other directly, but each such place is a workplace (which is a randomly chosen clique from Layer $2$), and the visitors of the store are connected to the employees, not to each other. We get complete bipartite graphs this way. This also means that stores have a key role at the beginning of the epidemic, if some of the employees get infected, they can infect many other people, but once they are all recovered from the disease, the stores do not have an important role anymore. 

\section{Epidemic spread and role of the parameters}

\label{sec:epidemic}

Now we have constructed all four layers of the underlying graph. Notice that randomness in the graph comes from the distribution of the households at the nodes of the grid in the geometric structure, and from the steps when we form schools, classes, workplaces, and nursing homes, by choosing groups of the same size uniformly at random. On the other hand, the connections in the different layers are not equally strong, hence we will use different infection parameters. In particular, we will use the  set of parameters listed in Table \ref{tab:parameters}. 

\begin{table}
\centering
\begin{tabular}{ccccc}
    layer & group & size & infection rate & value of $\tau$  \\ \hline 
    layer 1 & household & random  & $\tau_{\rm h}$ & $1$ \\ \hline layer 2 & class & $8 + 1 + 1$ & $\tau_{\rm cl}$ & $1$\\  & school & $200$ & $\tau_{\rm cl}\cdot \tau_{\rm sch}$ & $0.03$\\
  &  workplace & $10$ & $\tau_{\rm wp}$ & 1 \\
   & nursing home & $0+1+10$ & $\tau_{\rm wp}$ & $1$ \\
   \hline 
   layer 3 & neighbors & random & $\tau_{\rm g}$ & $0.1$ \\
   \hline layer 4 & store & 10 + 200  & $\tau_{\rm st}$  & $0.1$
\end{tabular}
\caption{The list of size and infection parameters for the four layers of our random graph; the last column contains the value of the infection rate in the baseline scenario (the recovery rate is $\gamma=1$ fixed).}
\label{tab:parameters}
\end{table}

Given the graph and the infection parameter corresponding to each edge of the graph, we run a continuous-time SIR process to model epidemic spread (see e.g.\ \cite{simon}). That is, the individual is always in one of the states susceptible (S), infected (I), or recovered (R). Recovered individuals remain in state R for the rest of the process. Infected individuals recover after a random time span, which is exponentially distributed with expectation $1/\gamma=1$. For each SI edge $e$ (an edge with an endpoint in state S and the other in state I), we associate a random variable $X_e$ with exponential distribution, with expectation $1/\tau$, with the infection parameter $\tau$ corresponding to that edge. The infection parameter depends on the layer containing the edge, and these values are given by  \autoref{tab:parameters} (the discussion on the choice of parameters can be found in the next subsection).  If edge $e$ becomes an SI edge at time $T_e$, and it is still an SI edge at time $T_e+X_e$, then its susceptible endpoint becomes infected at time $T_e+X_e$. All exponential random variables that we use for the recovery of infected vertices or the infection on SI edges are independent. 

In our computer simulations, the SIR process is performed by using the Gillespie algorithm, in the form described in \cite{simon}. The size of the graph, that is, the number of individuals (vertices) was chosen to be equal to $4000$ (in \cite{fourlayer} we observed that the proportion of infected vertices had a similar behavior for graphs of different sizes in a very similar model to the current one, so now we only used this smaller value, corresponding to a smaller city). 

\subsection{Choice of parameters}

\label{sec:parameters}

In this section, we will compare a few characteristics of the epidemic curve to real data from the Covid-19 (SARS-CoV-2) pandemic. However, since our main goal is to understand the effect of the structure of the random graph model, and not to examine the real-world effects (e.g.\ lack of testing capacity, selection bias, reporting recovery late), we do not fit our epidemic curves to real ones directly. 

As it was described in  \autoref{ref:graph}, the structure of the graph was constructed such that it approximately meets the data from the census  and from the survey \cite{maszk} (recall  \autoref{sec:data}). Of course, the choice of the infection parameters (the weight of the different types of contacts) is also very important, hence we need further points where we can check that our model is realistic. 

The first approach is to calculate the basic reproduction number $R_0$, that is, the expected number of people infected by a single individual in state I at the beginning when every other individual is susceptible. 

\begin{itemize}

\item We start with Layer 1, that is, the layer of households. Here the question is the following: if we choose a person uniformly at random, what is the expected number of people who live in the same household? Based on \autoref{tab:household}, this value is $2.13$. Since the infection rate and the recovery rate are the same within a household, the probability that an infected person transmits the disease to a given household member is equal to $1/2$. Hence the expected number of infections caused by a single individual within a household is $1.06$. 

\item Let us continue with layer 3, as age groups do not have a role here. Based on  \autoref{tab:household}, we can calculate that the expected size of a randomly chosen household is $2.43$. Hence we have to add the following term:
\[\frac{\tau_{\rm g}}{\tau_{\rm g}+\gamma}\cdot 2.43=0.26.\]

\item Now we turn to Layers 2 and 4, where we also have to take into account the age group of the person who is infectious at the moment.  In  particular, in the first case, we assume that an individual of age group $1$
gets infected. The expected number of people of age group $1$ who get infected (if all other people are in state $S$) is as follows: 
\[\frac{\tau_{\rm cls}}{\tau_{\rm cls}+\gamma}\cdot 7+\frac{\tau_{\rm cls}\cdot \tau_{\rm sch}}{\tau_{\rm cls}\cdot \tau_{\rm sch}+\gamma}\cdot 192=9.09.\] 

Here we used that each person of age group $1$ has $7$ classmates according to  \autoref{tab:parameters}, and there are $192$ other students in the same school. Since infection and recovery are described by independent, exponentially distributed random variables of parameter $\tau$ and $\gamma$, and infection occurs if and only if the first one is smaller than the latter, a well-known formula gives the probability of infection. 

\item The expected number of individuals in age group $2$ who are infected by an infected young person is as follows:  
\[\frac{\tau_{\rm cls}}{\tau_{\rm cls}+\gamma}+\frac{\tau_{\rm cls}\cdot \tau_{\rm sch}}{\tau_{\rm cls}\cdot \tau_{\rm sch}+\gamma}\cdot 24+\frac{\tau_{\rm st}}{\tau_{\rm st}+\gamma}\cdot 10=2.11.\]

Here the first term corresponds to the adult member of the class of this young person, the second term corresponds to the adult members of the other classes in the same school. Then comes the staff of the store, $10$ people; recall that all individuals attend exactly one store. 

\item The role of adults and elderly people is symmetric in the school layer, as there is exactly one adult and exactly one elderly person in each class. Hence the expected number of individuals in age group $3$ who are infected by an infected individual of age group $2$ is given by the same formula as above, without the term corresponding to the stores: 
\[\frac{\tau_{\rm cls}}{\tau_{\rm cls}+\gamma}+\frac{\tau_{\rm cls}\cdot \tau_{\rm sch}}{\tau_{\rm cls}\cdot \tau_{\rm sch}+\gamma}\cdot 24=1.2.\]

\item Now let us calculate the average number of individuals in age group $1$ who are infected by a randomly chosen infected adult. Here we also have to take into account that some of the adults are teachers, and some of them work in stores; however, for the sake of simplicity, we suppose that the infection affects all adults with the same probability. More precisely, if the total population was $4000$, then the number of stores is $4000/200=20$, so the number of adults working in stores is equal to $200$. The number of young people in a store is $200$ multiplied by the proportion of young people, which is $27.2\%$ by \autoref{tab:household}, which is approximately $54$. On the other hand, the total number of adults is  approximately $4000\cdot 55.5\%=2220$. The number of young people is $4000\cdot 27.2\%=1088$, so the number of classes in schools is $1088/8=136$. Therefore the expected number of young people infected by an adult can be calculated as follows (notice that there are only proportions in the formula, which do not depend on the total size of the population):
\[\frac{\tau_{\rm cls}}{\tau_{\rm cls}+\gamma}\cdot \frac{136}{2220}\cdot 8+\frac{\tau_{\rm cls}\cdot \tau_{\rm sch}}{\tau_{\rm cls}\cdot \tau_{\rm sch}+\gamma}\cdot\frac{136}{2200}\cdot 192+\frac{\tau_{\rm st}}{\tau_{\rm st}+\gamma}\cdot \frac{200}{2220} \cdot 54=0.91.\]

\item Now we continue with the expected number of adults infected by a randomly chosen infected adult. Adults meet each other at workplaces, stores, and in schools, if they belong to the same school, but they are in different classes. Hence, similarly to the calculations above, we obtain that the expected number of people infected by a randomly chosen adult who is infected is as follows: 
\[\frac{\tau_{\rm wp}}{\tau_{\rm wp}+\gamma}\cdot 9 + \frac{\tau_{\rm st}}{\tau_{\rm st}+\gamma}\cdot \frac{200}{2220} \cdot 111+\frac{\tau_{\rm cls}\cdot \tau_{\rm sch}}{\tau_{\rm cls}\cdot \tau_{\rm sch}+\gamma}\cdot\frac{136}{2200}\cdot 24=5.13.\]

\item Going further, adults meet elderly people at schools (within the same class and in different classes as well), stores, and nursing homes. In this last case, each adult meets $10$ people from age group $3$ according to \autoref{tab:parameters}. Hence we obtain that the expected number of infected people in age group $3$ caused by a randomly chosen adult who is infected is as follows: 

\[\frac{\tau_{\rm cls}}{\tau_{\rm cls}+\gamma}\cdot \frac{136}{2220}+\frac{\tau_{\rm cls}\cdot \tau_{\rm sch}}{\tau_{\rm cls}\cdot \tau_{\rm sch}+\gamma}\cdot\frac{136}{2200}\cdot 24+\frac{\tau_{\rm st}}{\tau_{\rm st}+\gamma}\cdot \frac{200}{2220} \cdot 35+\frac{\tau_{\rm wp}}{\tau_{\rm wp}+\gamma}\cdot \frac{56}{2200}\cdot 10=0.58.\]

Here we used that  $80\%$ of age group $3$ are in a nursing home, that is, $4000\cdot 17.3\%\cdot 80\%=554$ people. Hence the number of nursing homes is $56$, this is equal to the number of adults working in nursing homes, and we had $2200$ adults altogether.

\item Now let us calculate the expected number of young people infected by a randomly chosen individual of age group $3$ who is infectious. Here the only places that matter are schools. As we have already calculated, there are $136$ school classes, so $136$ elderly people work in schools. The total number of elderly people in our model is $4000\cdot 17.3\%=692$.   Hence the corresponding term in the basic reproduction number will be as follows: 
\[\frac{\tau_{\rm cls}}{\tau_{\rm cls}+\gamma}\cdot \frac{136}{692}\cdot 8+\frac{\tau_{\rm cls}\cdot \tau_{\rm sch}}{\tau_{\rm cls}\cdot \tau_{\rm sch}+\gamma}\cdot\frac{136}{692}\cdot 192=1.89.\]

\item Elderly people meet adults at schools, stores, and nursing homes as well. Hence the expected number of adults infected by a randomly chosen elderly person is given by 
\[\frac{\tau_{\rm cls}}{\tau_{\rm cls}+\gamma}\cdot \frac{136}{692}+\frac{\tau_{\rm cls}\cdot \tau_{\rm sch}}{\tau_{\rm cls}\cdot \tau_{\rm sch}+\gamma}\cdot\frac{136}{692}\cdot 24+\frac{\tau_{\rm st}}{\tau_{\rm st}+\gamma}\cdot 10+\frac{\tau_{\rm wp}}{\tau_{\rm wp}+\gamma}\cdot 0.8=2.04.\]

\item Finally, elderly people meet each other at schools in different classes, and in the  nursing homes. There were $136$ teachers, and   $80\%$ of the other $556$ elderly people are in nursing homes. Hence we get the following term: 
\[\frac{\tau_{\rm cls}\cdot \tau_{\rm sch}}{\tau_{\rm cls}\cdot \tau_{\rm sch}+\gamma}\cdot\frac{136}{692}\cdot 24+\frac{\tau_{\rm wp}}{\tau_{\rm wp}+\gamma}\cdot 9\cdot \frac{556}{692} \cdot 0.8=3.03.\]

\end{itemize}

A combination of these will be $R_0$. For the sake of simplicity, we assume that each person gets infected with the same probability. This means that we have to combine the values above with the proportion of the different age groups, in the layers where the number of infected people depended on the age group. Hence we obtain that an approximation of the basic reproduction number is 
\[R_0=1.06+0.26+0.27\cdot(9.09+2.11 + 1.2)+0.56\cdot(0.91 + 5.13+0.58 )+0.17\cdot(1.89 + 2.04+ 3.03 )=9.59.\]
 We did not find reliable estimates of the basic reproduction number in Hungary for Covid-19 (on the estimate of the effective reproduction number see \cite{effective}, where the authors find that the value was between $0.5$ and $1.5$ between April 2020 and January 2021, but this was observed under strict social distancing rules). For comparison, in Italy, the value of the basic reproduction number $R_0$ was between $2.4$ and $3.1$ in Spring 2020, during the first wave of the pandemic \cite{italy}. Our value is significantly larger. A possible reason for the difference is that not all infections are reported (either due to infection without symptoms, or lack of testing capacity), and this does not seem to be handled in \cite{italy}, so the basic reproduction number might be larger than what is observed there.  
 On the other hand, when we modified the parameters in order to get a smaller $R_0$, vaccination (see Section \ref{sec:vaccine}) suddenly stopped the spread of the epidemic, which does not seem realistic (the number of people vaccinated per day was chosen to match the proportion of people vaccinated in Hungary \footnote{\url{https://atlo.team/koronamonitor-reszletesadatok/}}). Furthermore, due to the small cliques in our model, the number of available susceptible people in the neighborhood of an infected person decreases very quickly, so the adequate reproduction number must be much smaller than the basic reproduction number, and hence the larger value of  $R_0$ might be appropriate. 
 
\begin{figure}
\begin{minipage}{0.5\textwidth}
\centering
\includegraphics[scale=0.4]{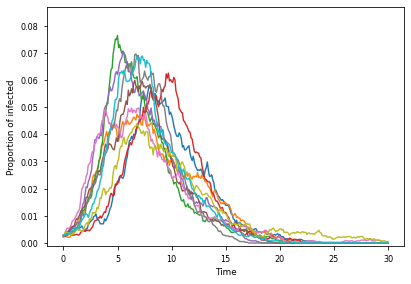}
\end{minipage}
\begin{minipage}{0.5\textwidth}
\centering
\includegraphics[scale=0.4]{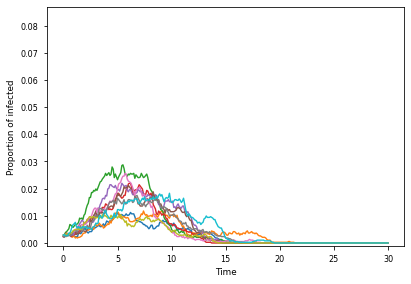}
\end{minipage}
\caption{Ten epidemic curves with the basic parameter setup given in \autoref{tab:parameters}, without vaccination (left), and with vaccination (right) }
\label{fig:tencurves}
\end{figure}

Another possibility to check whether our setup is realistic is based on the computer simulations themselves. In  \autoref{fig:tencurves}, we can see ten epidemic curves from the model with the values given in  \autoref{tab:parameters}, without and with vaccination. A detailed description of vaccination that we used in the computer simulations will be given in  \autoref{sec:vaccine}, here we only say that in this case vaccinated people were chosen randomly, independently of age or their position in the graph. As we can see, the maximal proportion of infected people is between $0.06-0.08$ and $0.01-0.03$ in these cases, which more or less corresponds to the observed data from Hungary (since we used census data from Hungary, it seems reasonable to use the epidemic data from the same country). In particular, according to \footnote{\url{https://atlo.team/koronamonitor-reszletesadatok/}}, the maximum number of infected people was around $270000$ in April 2021, while the population of the country was $9.7$ million by that time.  Hence the maximal proportion of infected people was $0.028$. This was observed in a period with certain social distancing rules, and with a small proportion of the population already vaccinated, and, of course, not all infections were reported. Hence we can say that the order of the maximal proportion of infected people on  \autoref{fig:tencurves} is realistic.

\begin{figure}[h]
\centering
\includegraphics[scale=0.4]{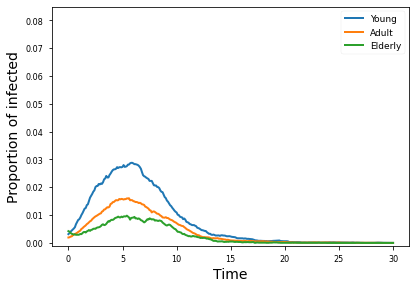}
\caption{Epidemic curves for the three age groups with vaccination}
\label{fig:agegroups}
\end{figure}

The observation of infected people is always less reliable than the data of the number of people in hospitals (as ill people do not necessarily go to official tests, but, during the covid-19 pandemic, all patients were tested upon their arrival at the hospital, independently of their disease or symptoms, so we can assume that infections were found with a much larger chance). \autoref{fig:agegroups} shows the average proportion of infected individuals in each age group. Each curve represents the average of 30 simulations.  We can see that approximately $1 \%$ of age group $3$  is infected at the peak of the epidemic wave \footnote{\url{https://atlo.team/koronamonitor-reszletesadatok/}}. On the other hand, in Hungary, the maximum number of hospitalized people was around $12000$. By assuming that most of them were elderly people, this would mean that people from age group $3$ are hospitalized with a probability of approximately $12000/(9700000\cdot 0.173)=0.72\%$, which is close to the values observed on  \autoref{fig:agegroups}.

\subsection{Parameter sensitivity}

In this subsection, we examine the effect of the different parameters on the spread of the epidemic process. These simulations contain information on the parameters which are the most important to estimate precisely if we wanted to fit the model to real data. On the other hand, the results are also useful from the point of view of controlling the epidemic: decision-makers can change the values of certain parameters through various actions. For example,  the case $\tau_{\rm cl}=0$ corresponds to the case when schools are closed. Hence our results also contain information on the effectiveness of the different social distancing strategies. 

\begin{figure}[h]
\centering
\includegraphics[scale=0.4]{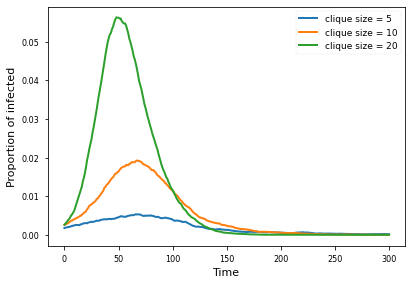}
\caption{Proportion of infected with different class  sizes}
\label{fig:class}
\end{figure}

As we could see in \autoref{sec:parameters} and also from the contact matrices \eqref{eq:contactc}, the most important components seem to be the schools in our model. Hence we are first interested in the effect of changing the sizes of the classes.  On  \autoref{fig:class}, we see that the maximal number of infected people and the total number of infected people are both  significantly different if we use different class sizes instead of the value 8 that we used (teachers are not counted in the class size, recall  \autoref{tab:parameters}). However, based on the results, we may assume that a smaller change in the class size would not cause a significant difference, so it is appropriate if we use a value that is close to the observed values in the contact matrix \eqref{eq:contacto}.

\begin{figure}[h]
\begin{minipage}{0.5\textwidth}
\centering
\includegraphics[scale=0.4]{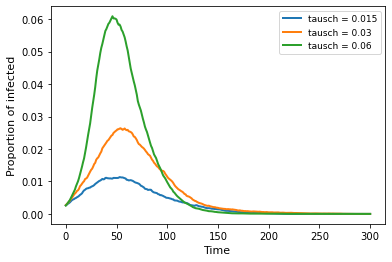}
\end{minipage}
\begin{minipage}{0.5\textwidth}
\centering
\includegraphics[scale=0.4]{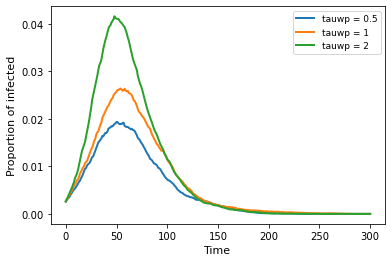}
\end{minipage}
\caption{Proportion of infected individuals with different strengths of connections between classes (left) and  with different strengths of connections in the workplaces (right)}
\label{fig:tausch}
\end{figure}

On  \autoref{fig:tausch}, we can compare the effect of changing the infection parameter $\tau_{\rm sch}$ corresponding to schools, and the infection parameter within workplaces, $\tau_{\rm wp}$ (recall  \autoref{tab:parameters}). Each curve is obtained as  the average of the results from $30$ independent simulations.  In both cases, the basic scenario corresponds to the case in the middle (orange). We can see here again that schools have a more significant role (recall from  \autoref{sec:parameters} that the term in $R_0$ corresponding to the schools was the largest one among all the terms), as this infection parameter has a stronger effect on the epidemic spread. Since $\tau_{\rm sch}$ belongs to the connections between members of different classes, comparing  \autoref{fig:class} and  \autoref{fig:tausch} we conclude that reducing the frequency of contacts between classes can be almost as effective as reducing the class sizes if our goal is to reduce the size of the epidemic wave.   

\section{Comparison of vaccination strategies} 

\label{sec:vaccine}

In this section, our main goal is to compare different vaccination strategies which are based on age groups and also use the structure of the underlying graph. For example, it is interesting to see whether it is worth taking into account the clique structure of society when we choose the order of vaccination of the individuals. 

We have examined various vaccination strategies which all satisfy the following properties. The parameters were chosen such that we obtain a realistic scenario in the basic case where people were vaccinated in a uniformly random order. This can be seen in  \autoref{fig:tencurves}, and in  \autoref{sec:parameters} we could see that this indeed leads to trajectories that match the order of real  epidemic curves. 

\begin{enumerate}
\item We assume that, similarly to seasonal influenza, we have the possibility to vaccinate a given  proportion  of the population in advance, before infections start to spread. More precisely, $V_0=420$ people ($10.5\%$ of the population) were vaccinated by day $0$, that is, at the time when we have chosen the first infected individuals and the process started. 
\item From day $0$, in the first period, $30$ people are chosen for vaccination on each day. This matches the proportion of people vaccinated per day in Hungary \footnote{\url{https://atlo.team/koronamonitor-reszletesadatok/}}. On the other hand, in our model, as time goes on, the number of people who are willing to get the vaccine decreases. The final proportion of vaccinated people is around $60\%$. This can be seen in \autoref{fig:oltarany}.

\begin{figure}[h]
\centering
\includegraphics[scale=0.4]{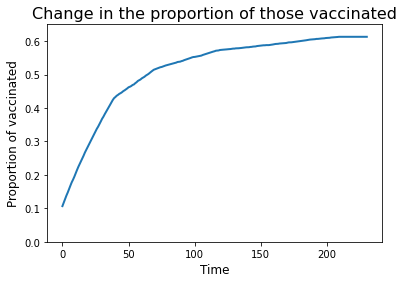}
\caption{Changes in the proportion of vaccinated people over time}
\label{fig:oltarany}
\end{figure}

\item Within a given group, depending on the actual vaccination strategy (e.g.\ elderly people or the members of a given school), the vaccinated people are chosen uniformly at random with replacement. When someone is chosen again, then nothing happens, this also means that the number of vaccinated people is somewhat less than the numbers given above. This might correspond to the fact that the people who have an appointment might not get vaccinated due to various reasons. 
\item When an individual receives a vaccination, they develop immunity to the virus, which gradually strengthens for two weeks. This immunity is represented by a numerical value ranging between 0 and 1. Let us denote this value by $p$, which signifies the probability that the person remains uninfected if exposed to the virus in a Gillespie simulation. If an individual has been targeted to the virus, they have two possible outcomes: they can either become infected or "recover," which implies they are no longer susceptible to reinfection.
\end{enumerate}

Within this framework, the first goal was to understand the role of age. We compared vaccination strategies in which (a) young (b) adult (c) elderly people were vaccinated first. More precisely, in each case, among the $V_0$ people who are vaccinated before the epidemic starts, have a given age distribution. For strategy "young", the distribution is $(80\%, 10\%, 10\%)$, that is, the majority of the people who get the vaccine in advance are from age group $1$. Similarly, for strategy "adult", we have age distribution  $(10\%, 80\%, 10\%)$, and for strategy "elderly", we have $(10\%, 10\%, 80\%)$. From day 1, we continue vaccination such that this distribution remains valid until $60\%$ of a certain age group gets vaccinated. Since this is the proportion of vaccinated people in general, we stop choosing people from a group when $60\%$ of the members are already vaccinated.

\begin{figure}[h]
\centering
\includegraphics[scale=0.4]{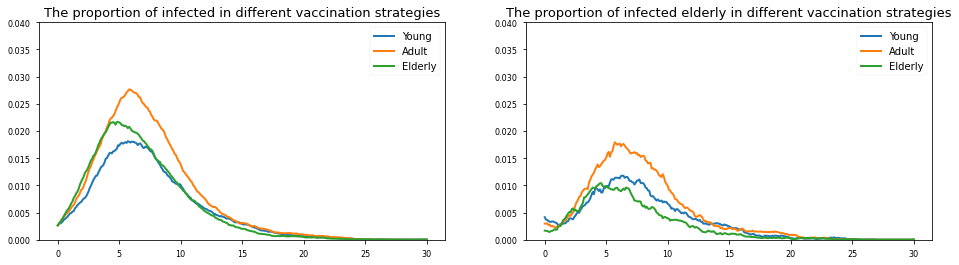}
\caption{Comparison of the pre-vaccination of young people, adults, and the elderly; the proportion of infected people in the total population (left) and within the age group 3, elderly people (right)}
\label{fig:strategies}
\end{figure}

On \autoref{fig:strategies}, we can see the proportion of infected and the proportion of infected elderly with these different vaccination strategies. In both cases, each curve is the average of $30$ independent simulations with the same set of parameters and vaccination strategy. The second quantity is  important for reducing mortality, and also from the point of view of the capacity of hospitals, as for most infections, it is the elderly people who have a significantly higher chance to be affected by the most serious consequences.  The results show that vaccinating young people first is the most effective if our goal is to reduce the number of infected people in the total population. This corresponds to the fact that, in our model, young people typically infect the most people (recall the calculations of the basic reproduction number $R_0$ from \autoref{sec:parameters}). However, if we take into account that elderly people have a higher death rate due to the virus, we can see that vaccinating elderly people is the best choice, although the strategy "young", which was the best in general, is almost as good as this one.

\begin{figure}[h]
\centering
\includegraphics[scale=0.4]{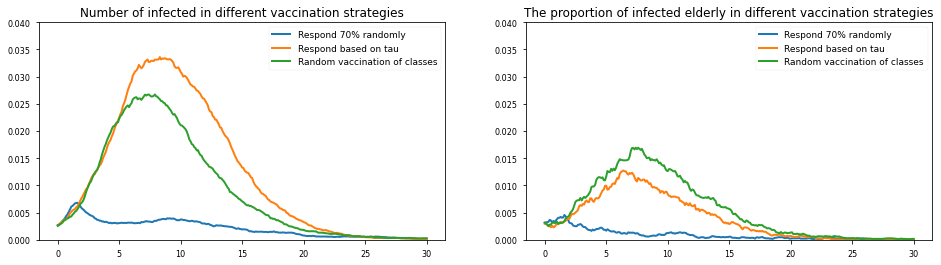}
\caption{Comparison of different vaccination strategies: respond 70\% randomly, respond based on $\tau$, and vaccinate classes randomly. }
\label{fig:agegroups2}
\end{figure}

By considering a reactive approach which is heavily based on the community structure of the graph, we implement a highly effective vaccination strategy. In this scenario, when an individual becomes infected, all the individuals they have been in contact with are promptly prioritized in the vaccination queue (notice that this makes it necessary to find the infected individuals quickly and precisely). This proactive measure significantly enhances the vaccination campaign's overall effectiveness as seen in \autoref{fig:agegroups2}. The three different strategies are: the contact will be in the next vaccination with probability $p = 0.7$; the contact will be in the next vaccination if the contact was strong ($\tau = 1$); or we vaccinate classes randomly (that is, the members of school classes and workplaces are vaccinated at the same time, and we always choose the next group uniformly at random).

Results suggest that if we take into account the contacts with a smaller infection rate, for example, relationships within a school, then we can prevent the rapid spread of the virus. Vaccinating whole classes randomly has basically the same effect in this simulation as vaccinating randomly (recall \autoref{fig:tencurves}, where we can see what happens if vaccinated individuals are chosen uniformly at random, independently of their age or position in the graph). However, targeting the communities of the infected individual proves to be considerably more effective in controlling the spread of infections.

\section{Conclusions}

Our main goal was to understand the role of the community structure in the process of epidemic spread, based on a SIR process running on a multilayer random graph model. Our model consists of four layers, and we set up the main parameters such that the weight of edges between individuals belonging to a certain age group matches the number of contacts observed in a recent survey. In our graph, the degree distribution was rather homogeneous.   

In this model, when we studied the role of different parameters, we could see that lowering the size of the school classes or the weight of the edges between students who attend different classes in the same school can reduce the size of the epidemic wave significantly, hence these are effective social distancing strategies. As for vaccination, the most successful strategy that we found was to vaccinate the members of the communities in which we find an infected individual. Of course, it is not always realistic to detect infection immediately, so we also compared strategies that depend only on age and hence they are more realistic. As it was expected, we found that if we focus on the protection of elderly people, then vaccinating them first is the best option. However, vaccinating young people first (that is, the people who have the largest number of contacts) is just slightly worse for the elderly, but it leads to fewer infections in the whole community, which also means that the healthcare system has larger capacities for those who are seriously ill. That is, we find that it is worth considering prioritizing people with more contacts, even if they are not the most vulnerable members of the population. In our model, all young people have approximately the same number of contacts; in real life, a possible strategy would be to vaccinate elderly people and young people with the highest number of contacts approximately at the same time. This conclusion is similar to the results of Knipl and R\"ost \cite[2011]{knipl}, who examined a compartment model with differential equations and a more detailed age and infection structure, but without any underlying graph. 

As for further directions and possible questions, it is important to mention that our random graph model could be easily formed into a weighted hypergraph model, where the communities are represented by hyperedges.  Still, in our case, infection spreads from one single individual to another. Similar questions about the role of the communities could be examined for opinion spread when several members in a group can "infect" other members with their opinion, that is, the spreading process is also based on interactions of more than two individuals. In this direction, we refer to the work of Landry and Restrepo \cite[2020]{hypergraph1}, Sun and Bianconi \cite[2021]{hypergraph2}, and the references therein.  

\subsection*{Acknowledgement}

This research was completed in the National Laboratory for Health Security RRF-2.3.1-
21-2022-00006 (Hungary).

\bibliographystyle{siam}
\bibliography{./main.bib}

\end{document}